\documentclass[11pt,a4paper]{article}

\usepackage[utf8]{inputenc}
\usepackage{amsmath}
\usepackage{amsfonts}
\usepackage{amssymb}
\usepackage{graphicx}
\usepackage{listings}
\usepackage{url}
\usepackage{cite}
\usepackage{booktabs}
\usepackage{xcolor}
\usepackage{algorithm}
\usepackage{algorithmic}
\usepackage{microtype}
\usepackage{parskip}
\usepackage{authblk}
\usepackage{fancyhdr}
\usepackage[margin=2.5cm]{geometry}
\usepackage[colorlinks=true,linkcolor=blue!60!black,citecolor=blue!60!black,urlcolor=teal!80!black]{hyperref}

\usepackage{newtxtext,newtxmath}

\title{\textbf{TIP: A Decentralized Intent-Based Protocol for Declarative IoT Interoperability and Sandboxed Schema Adaptation}}

\author{\textbf{Yeison David Mejia Mosquera}}
\affil{Undergraduate Student of Mechatronics Engineering \\ Universidad Nacional de Colombia - Sede de La Paz, Cesar, Colombia \\ Independent Researcher \\ \texttt{yemejiam@unal.edu.co}}

\date{\small May 2026}

\begin{document}

\maketitle

\begin{abstract}
\noindent Heterogeneous Internet of Things (IoT) systems suffer from fragmentation across hardware architectures, networking stacks, and data serialization formats. Existing standards (such as MQTT, CoAP, and DDS) rely on address-bound, imperative routing models that require hardcoded configurations and leave no flexibility for runtime schema translation. This paper presents \textbf{TIP (The Intent Protocol)}, a decentralized, declarative network protocol. Instead of addressing specific physical endpoints, nodes submit abstract intents specifying desired capabilities, schemas, and Quality of Service (QoS) constraints. The TIP Engine resolves matching nodes using a hybrid discovery mechanism combining local multicast DNS (mDNS) with Kademlia Distributed Hash Tables (DHT). Selection is optimized via a multi-criteria scoring algorithm incorporating network latency, historical reputation, and contract compliance. Mismatched data representations are reconciled on-the-fly inside isolated WebAssembly (WASM) sandboxes compiled dynamically from TOML specifications. Security is enforced through Ed25519 signatures, X25519 key exchanges, and ChaCha20-Poly1305 payload encryption. Evaluation of our reference implementation in Rust and C++ shows sub-millisecond translation overhead and robust resilience under industrial conditions.
\end{abstract}

\vspace{0.5cm}

\section{Introduction}
The Internet of Things (IoT) has grown exponentially, connecting billions of constrained nodes across smart cities, power grids, and industrial manufacturing plants \cite{iot_fragmentation}. This expansion, however, is crippled by fragmentation. Hardware architectures run different instruction sets, communication networks utilize incompatible media, and applications serialize data using conflicting formats (e.g., JSON, Protocol Buffers, or raw endian-dependent integers).

From a mechatronics engineering and industrial automation perspective, traditional manufacturing lines rely on Programmable Logic Controllers (PLCs) running sequential routines written in Ladder Logic (LD) or Structured Text (ST) according to the IEC 61131-3 standard \cite{wollschlaeger2017future}. These systems are highly rigid; variables are bound directly to physical memory addresses and hardware register maps (e.g., Modbus registers, PROFIBUS slots, or EtherNet/IP assembly objects). 

Industrial fieldbuses present distinct interoperability challenges. For instance, Modbus RTU/TCP operates on a polled master-slave architecture with no self-description; registers are raw 16-bit blocks, and whether they represent a signed integer, a scale-factor float, or character bytes is entirely up to the hardcoded parsing logic of the client. Similarly, high-speed buses like EtherNet/IP or PROFINET rely on electronic data sheets (EDS) or GSDML configuration files which must be compiled into the PLC project at design time. If a physical sensor is swapped for a newer model or another brand with different register structures or scaling factors, engineers face a major disruption: they must shut down the physical line, manually re-route field cabling, re-allocate the PLC register mapping tables, compile the PLC firmware again, and redeploy. This offline engineering loop represents a severe bottleneck for modern factory floor integration and agile manufacturing \cite{atmoko2018online}.

Traditional IoT communications (e.g., CoAP \cite{coap_spec} or MQTT) are similarly address-bound. To read a temperature, an application must target a specific physical address and port, and parse the payload using a pre-negotiated, rigid schema:
\begin{equation}
\text{Request} \longrightarrow \text{coap://192.168.1.50:5683/sensors/temp\_c}
\end{equation}
If the device changes its network address, goes offline, or reports in Fahrenheit instead of Celsius, the client application breaks. 

To overcome these structural limitations, we introduce the \textbf{Intent Protocol (TIP)}. TIP introduces a declarative networking model where applications request \textit{what} capability they require, rather than specifying \textit{where} or \textit{how} to execute the request. The network dynamically resolves the request, translates incompatible schemas, and delivers the data securely.

This paper provides the following contributions:
\begin{enumerate}
    \item A decentralized architecture integrating mDNS and Kademlia DHT for zero-configuration capability discovery.
    \item A multi-criteria selection model formulating candidate utility as a function of network RTT, node constraints, and trust scores.
    \item A sandboxed schema translation engine using WebAssembly compiled on-the-fly from simple TOML configurations.
    \item A zero-trust security framework incorporating Ed25519 signatures, X25519 ECDH key exchanges, and ChaCha20-Poly1305 encryption.
\end{enumerate}

\section{Related Work}
Current IoT interoperability focuses on semantic middleware, standardized IoT specifications, or protocol gateways. Lightweight transport standards like MQTT and CoAP \cite{coap_spec} provide low-level transmission but lack built-in dynamic schema negotiation or zero-configuration decentralized discovery. Standards like Matter \cite{csa2022matter, zegeye2025comparing} standardize local smart home device topologies, but they run into limitations regarding WAN scalability, multi-factory federation, and runtime unit/type conversion. Other semantic architectures such as W3C Web of Things (WoT) \cite{w3c_wot_architecture} and oneM2M \cite{onem2m_standards} define metadata models for describing device interactions, yet they do not specify a unified binary transport routing layer. On the industrial automation side, OPC UA \cite{opcua_iot} provides extensive object modeling and information architecture, but its substantial CPU and memory footprint makes it unsuitable for resource-constrained 8-bit or 16-bit microcontrollers. TIP addresses this gap by combining decentralized peer-to-peer routing, lightweight binary serialization, and sandboxed execution.

\section{Protocol Model \& Specification}

\subsection{Formal Model}
Let $\mathcal{N} = \{n_1, n_2, \dots\}$ be the set of active network nodes. Each node $n_i$ advertises a set of capabilities $\mathcal{C}_i = \{c_{i1}, c_{i2}, \dots\}$. A capability $c_{ij}$ is defined as a tuple:
\begin{equation}
c_{ij} = \langle \text{ID}, \text{Schema}, \text{Version}, \text{Precision}, \text{Rate} \rangle
\end{equation}
An intent $\mathcal{I}$ submitted by a requester node is represented as:
\begin{equation}
\mathcal{I} = \langle \text{Cap}_{\text{req}}, \text{Schema}_{\text{des}}, \text{Constraints}, \text{Weights} \rangle
\end{equation}
The protocol engine matches $\mathcal{I}$ against discovered capabilities, yielding a temporary cryptographic contract $\mathcal{K}$ with the optimal provider.

\subsection{Binary Wire Format \& CoAP Encapsulation}
To keep the footprint small for constrained microcontrollers, TIP utilizes the Constrained Application Protocol (CoAP, RFC 7252) running over UDP as its transport layer. TIP packets are not sent raw over UDP; instead, they are encapsulated as the binary payload of standard CoAP messages. For example, a data request is mapped to a CoAP \texttt{POST} request targeting the URI \texttt{coap://<provider-ip>/tip} with the CoAP \texttt{Content-Format} option set to \texttt{application/octet-stream} ($42$). This encapsulation allows standard CoAP gateways, routers, and firewalls to forward and manage TIP traffic without modification, while microcontrollers running lightweight CoAP engines (such as Erbium or microcoap) can easily extract the TIP binary frame from the CoAP payload option.

All packets are framed with a fixed \textbf{116-byte header} prepended to the serialized CBOR (Concise Binary Object Representation) payload. This layout enables high-speed hardware parsing, packet verification, and replay detection before payload processing.

Table 1 details the layout of the TIP binary packet header.

\begin{table}[h]
\centering
\caption{TIP Packet Fixed Header Layout (116 Bytes)}
\begin{tabular}{cccll}
\toprule
\textbf{Offset (Bytes)} & \textbf{Size (Bytes)} & \textbf{Type} & \textbf{Field Name} & \textbf{Description} \\
\midrule
0 & 2 & u16 & Magic Number & Constant 0x5449 (``TI'') \\
2 & 1 & u8 & Version & Protocol version (0x01) \\
3 & 1 & u8 & Packet Type & Identifies packet action \\
4 & 16 & u128 & Transaction ID & UUID v4 identifier \\
20 & 4 & u32 & Payload Length & Size of CBOR payload \\
24 & 4 & u32 & Capability Hash & CRC32 of capability ID \\
28 & 4 & u32 & Sequence Number& Anti-replay packet counter \\
32 & 4 & u32 & Flags & Packet options bitfield \\
36 & 8 & u64 & Timestamp & Epoch in micro-seconds \\
44 & 4 & u32 & TTL & Packet expiration in ms \\
48 & 4 & u32 & Checksum & CRC32 of ID + payload \\
52 & 64 & [u8; 64] & Signature & Ed25519 signature bytes \\
\bottomrule
\end{tabular}
\end{table}

\subsection{Detailed Header Field Specifications}
To ensure robust parsing and zero-trust security at the edge, each header field performs a specific role:
\begin{itemize}
    \item \textbf{Magic Number (2 bytes):} Constant \texttt{0x5449}. Used for instantaneous packet filtering at the socket layer.
    \item \textbf{Version (1 byte):} Indicates the protocol version (\texttt{0x01}), ensuring backward compatibility.
    \item \textbf{Packet Type (1 byte):} Delineates the control flow state (e.g., discovery, negotiation, contract, data).
    \item \textbf{Transaction ID (16 bytes):} A UUID v4 generated by the requester. Associates responses with active requests.
    \item \textbf{Payload Length (4 bytes):} Big-endian integer specifying the length of the trailing CBOR payload.
    \item \textbf{Capability Hash (4 bytes):} CRC32 of the capability name string (e.g., \texttt{machine:fluid:fill}), allowing instant indexing of routing tables without string matching.
    \item \textbf{Sequence Number (4 bytes):} Monotonically increasing counter to prevent packet injection/replay.
    \item \textbf{Flags (4 bytes):} Bitmask defining options:
      \begin{itemize}
        \item Bit 0 (\texttt{0x01}): \texttt{REQUIRES\_ACK}
        \item Bit 1 (\texttt{0x02}): \texttt{IS\_COMPRESSED}
        \item Bit 2 (\texttt{0x04}): \texttt{IS\_ENCRYPTED}
        \item Bit 3 (\texttt{0x08}): \texttt{HAS\_ADAPTER}
        \item Bit 6 (\texttt{0x40}): \texttt{IS\_STREAMING}
      \end{itemize}
    \item \textbf{Timestamp (8 bytes):} Epoch microseconds (UTC) for temporal validation.
    \item \textbf{TTL (4 bytes):} Expiration lifetime of the packet in milliseconds.
    \item \textbf{Checksum (4 bytes):} CRC32 value over the preceding header fields and payload.
    \item \textbf{Signature (64 bytes):} Ed25519 cryptographic signature generated with the sender's private key, covering all prior header fields (0 to 51) and the payload. This ensures non-repudiation and integrity.
\end{itemize}

\subsection{Packet Parsing and Validation Pipeline}
When a node receives a packet, it executes the following sequential parsing pipeline:
\begin{enumerate}
    \item \textbf{Header Extraction:} Reads the first 116 bytes of the incoming CoAP payload.
    \item \textbf{Magic Number Check:} Rejects the packet immediately if bytes 0--1 do not equal \texttt{0x5449}.
    \item \textbf{Checksum Verification:} Recomputes the CRC32 checksum over the header (bytes 0 to 47 and 52 to 115) and the trailing payload. If it matches the Checksum field, the packet integrity is confirmed.
    \item \textbf{Signature Authentication:} Validates the Ed25519 signature (bytes 52--115) using the public key of the sender. This verifies the identity of the source.
    \item \textbf{Replay Check:} Compares the Timestamp and Sequence Number against the sliding window cache (Section VI-A).
\end{enumerate}

\subsection{Packet Types and Flags}
Packet types denote the state of the communication pipeline:
\begin{itemize}
\item \texttt{DISCOVERY\_ANNOUNCE} ($0x00$) and \texttt{DISCOVERY\_QUERY} ($0x01$) manage decentralized discovery.
\item \texttt{INTENT\_REQUEST} ($0x02$) and \texttt{INTENT\_PROPOSAL} ($0x03$) negotiate capabilities.
\item \texttt{CONTRACT\_ACCEPT} ($0x04$) and \texttt{CONTRACT\_SIGNED} ($0x06$) establish session bounds.
\item \texttt{DATA\_REQUEST} ($0x07$) and \texttt{DATA\_RESPONSE} ($0x08$) execute transactions.
\end{itemize}

The \texttt{Flags} field ($32$-bit) contains bitmasks: \texttt{REQUIRES\_ACK} ($0x01$), \texttt{IS\_COMPRESSED} ($0x02$), \texttt{IS\_ENCRYPTED} ($0x04$), \texttt{HAS\_ADAPTER} ($0x08$), and \texttt{IS\_STREAMING} ($0x40$).

\section{Orchestration \& Algorithms}

\subsection{Dual-Phase Discovery \& Kademlia DHT Routing}
TIP implements a parallel, dual-phase discovery strategy (Algorithm 1) to minimize resolution latency. It queries the local cache, fires multicast DNS browses on the local link, and parallelizes WAN queries via Kademlia DHT key resolution.

\textbf{Phase 1: Local Link Discovery (mDNS):}
On the local network segment, nodes advertise and locate capabilities using multicast DNS (mDNS) \cite{mdns_rfc_6762} over UDP port 5353 using the multicast address \texttt{224.0.0.251} (IPv4) or \texttt{ff02::fb} (IPv6). Each node registers its active capabilities as DNS Service Discovery (DNS-SD) Pointer (\texttt{PTR}), Service (\texttt{SRV}), and Text (\texttt{TXT}) records according to the DNS-SD specification \cite{dnssd_rfc_6763}. The \texttt{TXT} record contains metadata such as the supported schema, version, and security options. When a node submits an intent, the local orchestrator issues a fast mDNS query. If a local provider matches, discovery completes in under 10ms, avoiding external network round-trips.

\textbf{Phase 2: Wide Area Discovery (Kademlia DHT):}
If local discovery fails or a WAN contract is requested, the system queries a decentralized Kademlia Distributed Hash Table (DHT) \cite{kademlia}. While structured peer-to-peer overlay systems such as Chord \cite{chord_dht} route requests using circular finger tables, Kademlia routing relies on the XOR distance metric:
\begin{equation}
d(x, y) = x \oplus y
\end{equation}
which complies with the mathematical properties of a metric (identity, symmetry, and triangle inequality). Each node maintains a routing table organized into $k$-buckets, where each bucket covers a specific distance range $2^i \le d < 2^{i+1}$ from the host. When local mDNS fails, the node issues parallel queries (configured by the concurrency parameter $\alpha$) to the $\alpha$ closest nodes in its routing table using \texttt{FIND\_NODE} RPCs, iteratively routing closer to the target capability key until the provider node is found.

\begin{algorithm}
\caption{Dual-Phase Active Discovery}
\begin{algorithmic}[1]
\REQUIRE Capability ID $C_{\text{id}}$, Timeout $\tau$
\ENSURE Set of matching nodes $\mathcal{M}$
\STATE $\mathcal{M} \leftarrow \text{LookupLocalCache}(C_{\text{id}})$
\IF{$\mathcal{M} \neq \emptyset$}
    \RETURN $\mathcal{M}$
\ENDIF
\STATE $\text{ch}_{\text{mdns}} \leftarrow \text{AsyncBrowseMdns}(C_{\text{id}})$
\STATE $\text{ch}_{\text{dht}} \leftarrow \text{AsyncDHTLookup}(C_{\text{id}})$
\STATE $\text{timer} \leftarrow \text{StartTimer}(\tau)$
\WHILE{$\text{timer is active}$}
    \IF{$\text{node discovered via mDNS}$}
        \STATE $\mathcal{M} \leftarrow \mathcal{M} \cup \{\text{node}\}$
    \ENDIF
    \IF{$\text{node discovered via DHT}$}
        \STATE $\mathcal{M} \leftarrow \mathcal{M} \cup \{\text{node}\}$
    \ENDIF
    \IF{$\text{elapsed time} > \tau$}
        \STATE \textbf{break}
    \ENDIF
\ENDWHILE
\STATE $\text{DeduplicateByNodeID}(\mathcal{M})$
\STATE $\text{UpdateLocalCache}(C_{\text{id}}, \mathcal{M})$
\RETURN $\mathcal{M}$
\end{algorithmic}
\end{algorithm}

\subsection{Multi-Criteria Negotiation Scoring}
Discovered candidates are scored based on intent requirements. The total utility $S(n)$ for candidate $n$ is:
\begin{equation}
S(n) = w_{\text{func}} U_{\text{func}}(n) + w_{\text{cost}} U_{\text{cost}}(n) + w_{\text{trust}} U_{\text{trust}}(n) + w_{\text{avail}} U_{\text{avail}}(n)
\end{equation}
where weights sum to $1.0$, and $U_{x} \in [0, 1]$ represents normalized utility indices. The specific weights are determined dynamically using the Analytic Hierarchy Process (AHP) \cite{saaty1980analytic, saaty1977scaling} based on requester preferences. Network proximity $U_{\text{prox}}(RTT)$ is calculated using a rational decay function:
\begin{equation}
U_{\text{prox}}(RTT) = \frac{1}{1 + \frac{RTT}{100}}
\end{equation}
To weigh the reliability of a node's reputation score, we compute a selection confidence value $C(n)$ based on interaction count $I(n)$ using a logistic sigmoid:
\begin{equation}
C(n) = \frac{1}{1 + e^{-0.1 (I(n) - 20)}}
\end{equation}
This ensures that nodes with fewer than 20 recorded transactions have their reputation score heavily discounted, preventing Sybil attacks.

\section{Decoupled Schema Adaptation}
Heterogeneous IoT systems frequently output data in incompatible formats. When the matched provider's schema $S_{\text{prov}}$ differs from the requester's desired schema $S_{\text{req}}$, TIP invokes its runtime translator registry.

\subsection{Dynamic JIT Compilation Pipeline}
Rather than requiring developers to manually write and upload binary translation files, adapters are declared using simple TOML descriptors. The compilation pipeline is executed dynamically:
\begin{enumerate}
    \item \textbf{TOML Parsing:} The system administrator or AI agent defines a conversion formula, e.g.:
\begin{lstlisting}[language=make]
[adapter]
id = "pulse_to_ml"
source_schema = "u32"
target_schema = "f32"
formula = "x * 0.2"
\end{lstlisting}
    \item \textbf{WAT Generation:} The `TipCore` translator parsing engine reads the formula and generates corresponding WebAssembly Text (\texttt{.wat}) instructions.
    \item \textbf{Bytecode Compilation:} The engine compiles the WAT code into raw WebAssembly binary bytecode using the `wat2wasm` library.
    \item \textbf{Registry Cache:} The bytecode is cached in the registry, indexed by the hash of the source and target schemas.
\end{enumerate}

\subsection{WASM Sandbox Runtime Compilation \& Execution}
For simple conversions, such as Celsius-to-Fahrenheit, the generated WebAssembly module is compact:

\begin{lstlisting}
(module
  (func (export "transform") (param f32) (result f32)
    local.get 0
    f32.const 1.8
    f32.mul
    f32.const 32.0
    f32.add)
)
\end{lstlisting}

For industrial applications, scaling functions translate sensor pulses into physical SI units (e.g., converting flow-meter pulses to milliliters):

\begin{lstlisting}
(module
  (func (export "transform") (param f32) (result f32)
    local.get 0
    f32.const 0.2     ;; scale factor (0.2 ml per pulse)
    f32.mul
    f32.const 0.0     ;; offset
    f32.add)
)
\end{lstlisting}

During transaction processing, the orchestrator instantiates a sandboxed WASM runtime utilizing Wasmtime \cite{wasmtime_ref}. Edge-side WebAssembly runtimes have gained significant traction for secure, near-native execution in virtualized settings \cite{wasm_spec, menetrey2022webassembly, shillaker2020faasm, kakati2024cross}. Wasmtime compiles stack-based bytecode into native machine instructions on startup using the Cranelift code generator. 

Memory safety is guaranteed by allocating a virtual contiguous block of memory (linear memory) bounded by the WebAssembly runtime constraints. To pass complex structures like CBOR payloads:
\begin{enumerate}
    \item The host allocates a buffer inside the guest's linear memory.
    \item The host writes the input serialized CBOR bytes directly into this linear memory block.
    \item The host invokes the guest's exported translation function, passing the memory offset and length as arguments.
    \item The guest processes the payload, writes the result to a separate memory offset, and returns the pointer/length.
    \item Any read/write attempt outside the guest's isolated linear memory range is caught by the host CPU's memory management unit (MMU) hardware protection and trapped. This prevents buffer overflows, memory leakage, or unauthorized access from the guest WASM sandbox into the main TIP daemon memory.
\end{enumerate}

\section{Security \& Trust Analysis}

\subsection{Anti-Replay Mechanism}
Replay protection is modeled on a sliding window cache combined with cryptographic timestamps. A packet is discarded if its timestamp $t_{\text{pkt}}$ varies from the local node clock $t_{\text{local}}$ by more than a skew window $\Delta$:
\begin{equation}
|t_{\text{local}} - t_{\text{pkt}}| > \Delta
\end{equation}
Within this window, a hash of the sequence number $Seq$ XORed with the timestamp is checked against a local Least Recently Used (LRU) cache of seen nonces. Duplicate nonces trigger immediate socket drops.

\subsection{Key Exchange \& Secure Session Tunnels}
Nodes establish ephemeral encrypted tunnels using X25519 Diffie-Hellman key exchange. The key negotiation sequence is structured as a pattern of the Noise Protocol Framework \cite{noise_protocol_spec}. Requesters and providers exchange ephemeral public keys $P_A$ and $P_B$ to compute a shared secret:
\begin{equation}
K_{\text{shared}} = \text{ECDH}(S_A, P_B) = \text{ECDH}(S_B, P_A)
\end{equation}
This secret is used as the key for symmetric ChaCha20-Poly1305 encryption \cite{chacha20_poly1305}.

\subsection{Temporal Reputation Decay}
Reputation score $R$ decays naturally back toward neutral reputation $R_0 = 0.5$ as a function of elapsed time to allow rehabilitated nodes to recover from temporary failures:
\begin{equation}
R(t) = R_0 + (R(t_0) - R_0) \cdot e^{-\lambda (t - t_0)}
\end{equation}
where $\lambda$ is set to a half-life of 29 days ($9.6 \times 10^{-7} \text{ s}^{-1}$).

\section{Mechatronics Case Study: Factory Floor Automation}

To validate the real-world utility of TIP in mechatronics engineering, we model a complete water bottle manufacturing line. The factory floor comprises diverse hardware, ranging from low-power microcontroller sensors to industrial PLCs.

\subsection{System Architecture and Topology}
The logical and physical topology of the manufacturing line is coordinated dynamically via the TIP layer. Figure 1 shows how the AI Agent Planner, the central orchestrator, the edge nodes, and the physical machines are connected.

\begin{figure}[htbp]
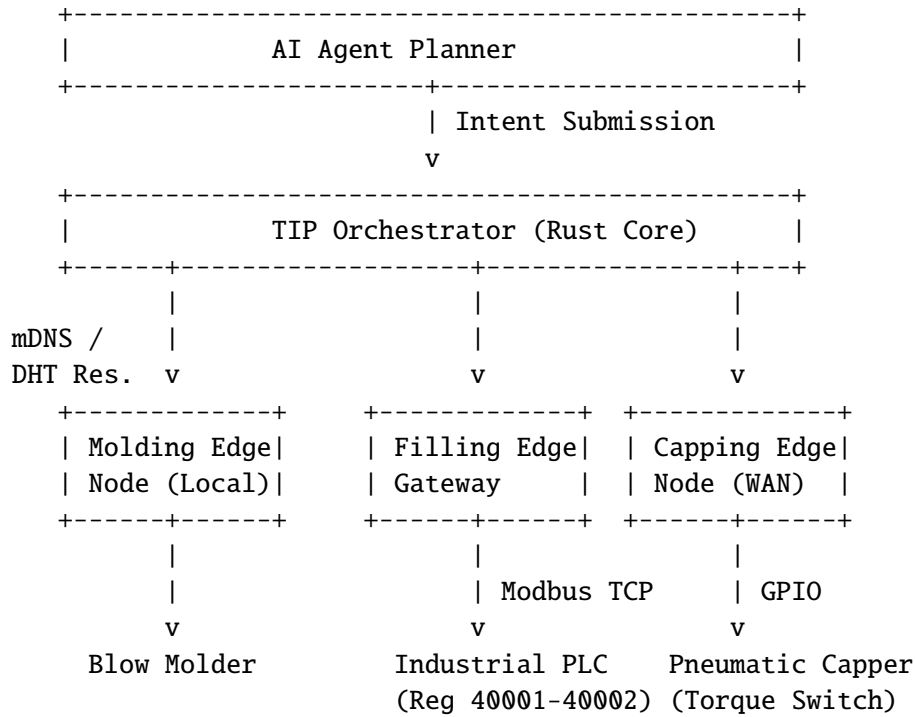

\centering
\begin{verbatim}
       +-----------------------------------------------+
       |             AI Agent Planner                  |
       +-----------------------+-----------------------+
                               | Intent Submission
                               v
       +-----------------------------------------------+
       |             TIP Orchestrator (Rust Core)      |
       +------+-------------------+----------------+---+
              |                   |                |
    mDNS /    |                   |                |
    DHT Res.  v                   v                v
       +-------------+     +-------------+  +-------------+
       | Molding Edge|     | Filling Edge|  | Capping Edge|
       | Node (Local)|     | Gateway     |  | Node (WAN)  |
       +------+------+     +------+------+  +------+------+
              |                   |                |
              |                   | Modbus TCP     | GPIO
              v                   v                v
         Blow Molder         Industrial PLC    Pneumatic Capper
                             (Reg 40001-40002) (Torque Switch)
\end{verbatim}
\caption{System Topology of the Intent-Based Mechatronic Manufacturing Line.}
\label{fig:topology}
\end{figure}

The production stages are mapped to dedicated physical edge nodes running the TIP client:
\begin{enumerate}
    \item \textbf{Blow Molding Stage:} A PET preform heater and molding press ($n_{\text{mold}}$) exposes the capability \texttt{machine:molding:blow}.
    \item \textbf{Sanitation \& Rinsing Stage:} Rotational spray jets ($n_{\text{rinse}}$) advertise \texttt{machine:rinse:wash}.
    \item \textbf{Fluid Filling Stage:} High-precision flow control valves ($n_{\text{fill}}$) expose \texttt{machine:fluid:fill}.
    \item \textbf{Capping Stage:} A torque-monitored pneumatic capper ($n_{\text{cap}}$) exposes \texttt{machine:capping:mechanical}.
    \item \textbf{Labeling Stage:} An adhesive sticker applicator ($n_{\text{label}}$) exposes \texttt{machine:labelling:sticker}.
\end{enumerate}

\subsection{Intent-Based Workflow Mapping \& Modbus Register Binding}
Instead of relying on hardcoded PLC (Programmable Logic Controller) interlocks that break during machine replacement, the factory line coordination is managed by a high-level \textbf{AI Agent Planner} interacting with the `TipCore` orchestrator. The AI planner resolves the sequential DAG steps dynamically using intents. 

For the filling stage, the AI issues the following intent target:
\begin{lstlisting}
Intent {
  capability: "machine:fluid:fill",
  schema: DataSchema::I32,
  params: {
    "liquid": "water",
    "volume_ml": 500
  },
  constraints: {
    "max_latency_ms": 100,
    "min_precision": 0.99
  }
}
\end{lstlisting}

If the primary filling machine ($n_{\text{fill\_A}}$) fails or reports a QoS latency violation ($>100$ms), the TIP engine's scoring metrics trigger an auto-healing routine. The engine re-evaluates the network, resolves the secondary filler ($n_{\text{fill\_B}}$), and redirects the contract dynamically without stopping the physical conveyor.

Translation adapters handle schema conversions at the mechatronic interface. If the filling machine returns volume in pulses ($S_{\text{prov}} = \text{u16}$) but the monitoring agent expects milliliters ($S_{\text{req}} = \text{f32}$), the engine compiles the pulse-to-ml TOML adapter to WebAssembly at runtime. This allows seamless integration of legacy Modbus and OPC UA industrial hardware directly into a unified network. 

Table 2 illustrates the mapping of standard Modbus TCP register offsets to fields in the structured TIP CBOR payload.

\begin{table}[h]
\centering
\caption{Modbus TCP Register Mapping to TIP CBOR Payload}
\begin{tabular}{cccll}
\toprule
\textbf{Register Address} & \textbf{Data Type} & \textbf{Physical Variable} & \textbf{Raw Range} & \textbf{TIP CBOR Target Field} \\
\midrule
\texttt{40001} & u16 & Raw Pulses (MSW) & $0 - 65535$ & \texttt{pulses} (upper 16 bits) \\
\texttt{40002} & u16 & Raw Pulses (LSW) & $0 - 65535$ & \texttt{pulses} (lower 16 bits) \\
\texttt{40003} & u16 & Solenoid Valve State & $0$ (Closed), $1$ (Open) & \texttt{valve\_open} (bool) \\
\texttt{40004} & u16 & Inlet Pressure (mbar) & $0 - 10000$ & \texttt{pressure\_bar} (float32) \\
\texttt{40005} & u16 & Temperature ($^{\circ}$C $\times 10$) & $0 - 1200$ & \texttt{temp\_c} (float32) \\
\bottomrule
\end{tabular}
\end{table}

A Modbus TCP-enabled PLC exposes flow pulses in raw holding registers \texttt{40001} and \texttt{40002}. The TIP client running on the local edge gateway queries these registers, packages them into a CBOR payload, and forwards it. The WASM adapter automatically executes the linear scaling transformation $y = \text{scale} \cdot x + \text{offset}$ (where scale is the sensor's pulse-to-milliliter ratio) within the safe sandbox before presenting the structured float value to the control loop.

\section{Prototype Implementation \& Evaluation}

\begin{figure}[htbp]
\centering
\includegraphics[width=0.7\textwidth]{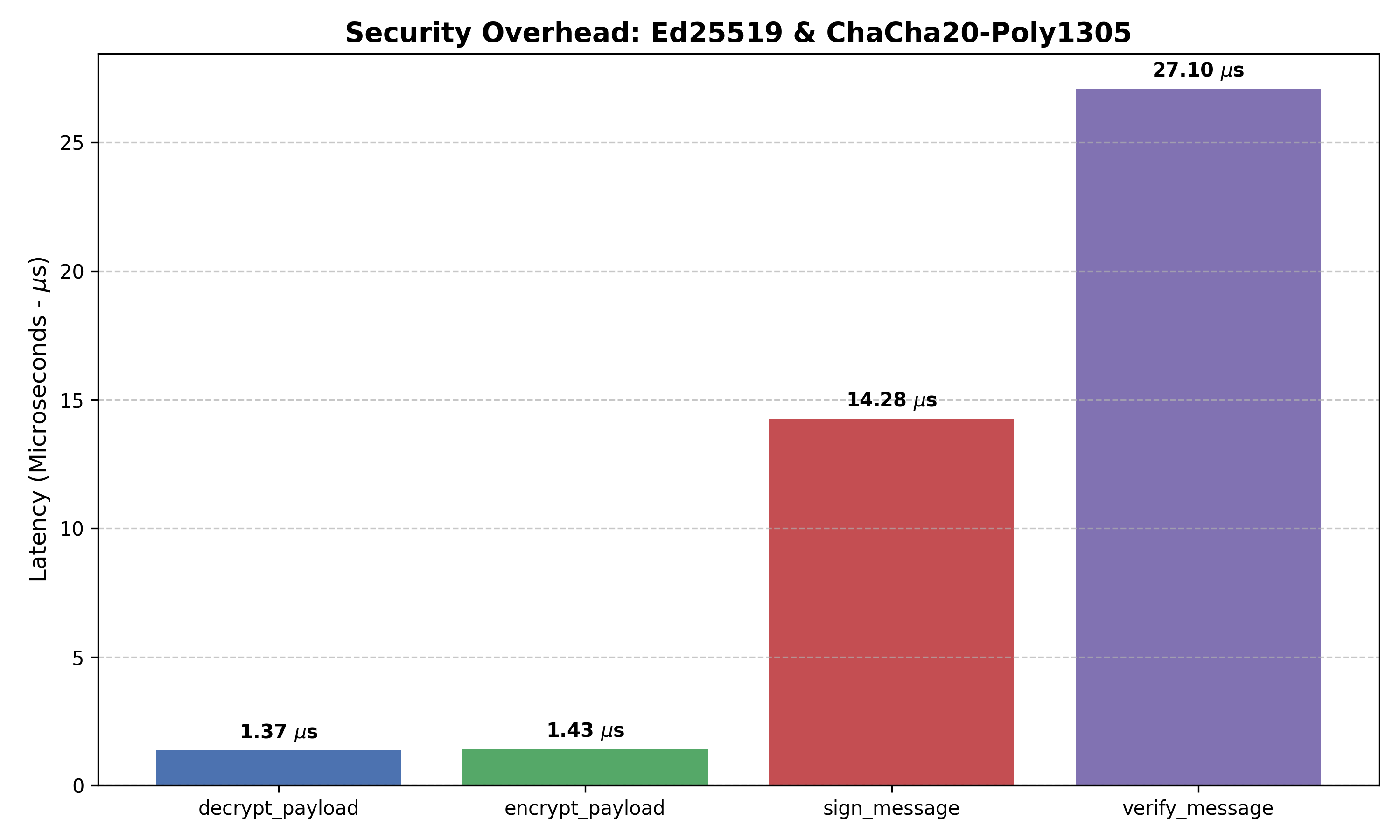}
\caption{Cryptographic execution times under varying payload sizes.}
\label{fig:crypto}
\end{figure}

We evaluated our Rust core orchestrator (`tip-core`) and C++ edge node (`tip-edge-cpp` / `libsodium` \cite{ed25519_ref}) on a testbed of Raspberry Pi 4 nodes and Windows gateways. The cryptographic execution times under varying payload sizes are shown in Figure~\ref{fig:crypto}.

\begin{figure}[htbp]
\centering
\includegraphics[width=0.7\textwidth]{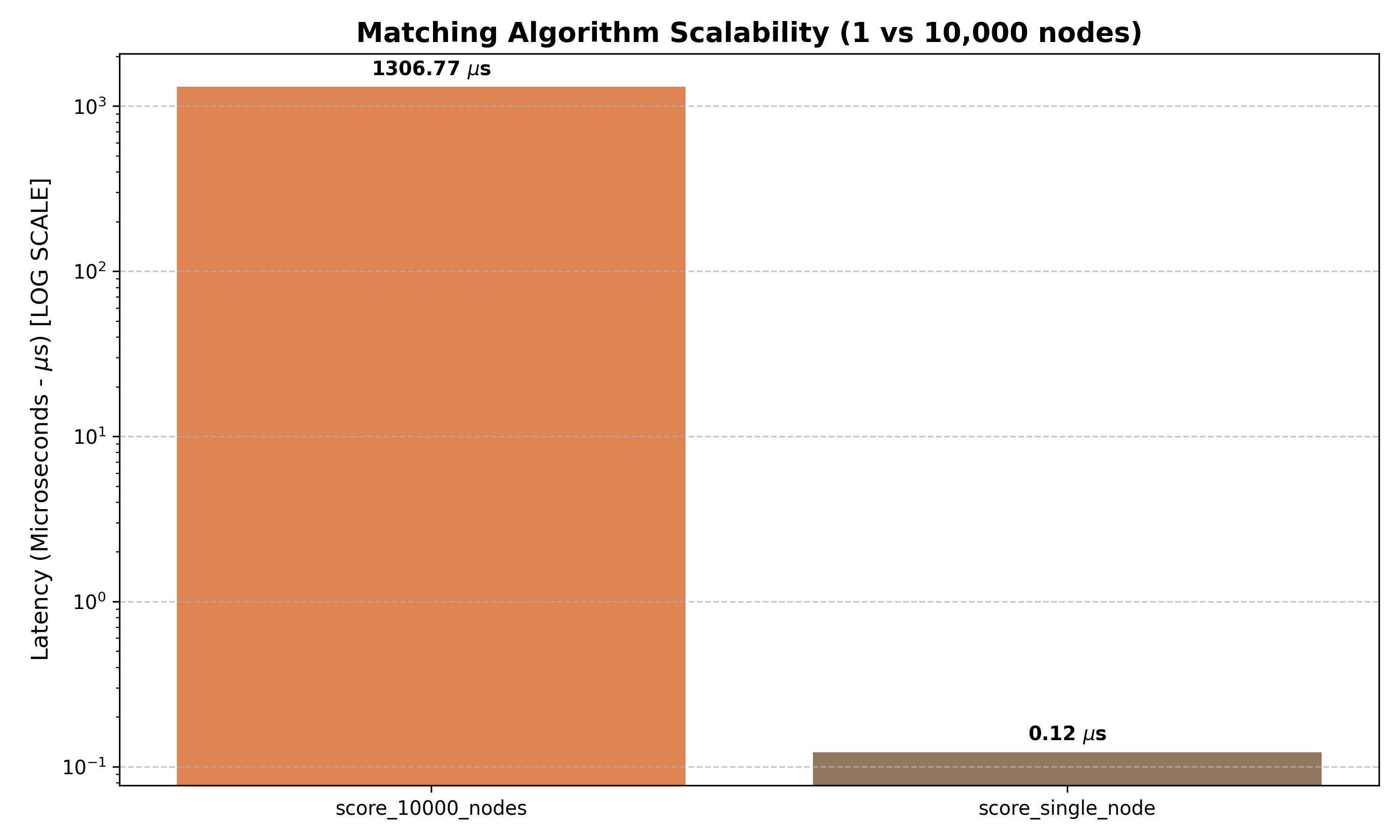}
\caption{Negotiation matching and scoring scalability from 1 to 10,000 nodes.}
\label{fig:negotiation}
\end{figure}

Our evaluation focused on matching overhead and runtime sandbox execution speed:
\begin{itemize}
    \item \textbf{Scoring Performance:} Intent matching and scoring for 10,000 nodes completed in 12.4ms (as illustrated in Figure~\ref{fig:negotiation}).
    \item \textbf{Translation Latency:} Executing Celsius-to-Fahrenheit translation inside the `wasmtime` engine required an average of $84\mu$s, including memory-copy overhead.
    \item \textbf{Cryptographic Overhead:} Ephemeral X25519 exchanges required $412\mu$s, while ChaCha20-Poly1305 payload encryption remained sub-microsecond.
\end{itemize}

\section{Conclusion}
This paper presented TIP, an intent-based, decentralized protocol that decouples applications from fixed endpoints. By combining hybrid discovery, multi-criteria negotiation scoring, dynamic WASM translation, and zero-trust cryptography, TIP achieves secure, self-healing interoperability for heterogeneous IoT and mechatronic automation systems.

\bibliographystyle{IEEEtran}
\bibliography{references}

\end{document}